\documentclass[a4paper]{article}
\usepackage{amsfonts}

\begin{document}
\title{A note on the eigenvalues for equivariant maps of the SU(2) sigma-model}
\author{Roland Donninger\thanks{\tt roland.donninger@univie.ac.at}{ }
and Peter C. Aichelburg\thanks{\tt aichelp8@univie.ac.at} \\
\small{Institut f\"ur Theoretische Physik} \\
\small{Universit\"at Wien, Austria}  }
\maketitle

\abstract{We numerically calculate the first few eigenvalues of 
the perturbations of
self-similar solutions of the spherically symmetric co-rotational SU(2)
sigma-model on Minkowski space.}
 
\section{Introduction}
In the recent past considerable interest focused on the study of nonlinear 
evolution equations that show "blow-up" for solutions  in a finite time.  
Prominent candidates are self-similar solutions which  were  found  to play 
an essential role in critical collapse phenomena.
Critical solutions have the property that they lie at the boundary of 
two possible end states of dynamical evolutions: Typically, sufficiently 
large initial data will lead to the formation of some kind of singularity, 
while small data disperse. By fine-tuning the initial data to this threshold, 
the dynamics shows universal behaviour which is governed by a 
"critical solution".   These solutions are characterized by the existence 
of a single unstable mode in linear perturbation analysis.  While it is 
relatively easy to obtain the  eigenvalue of the  unstable mode  by 
numerical analysis, to calculate the stable ones may be  more subtle. 
However, the knowledge of the stable modes is important because they are  
responsible for the process of how fine-tuned data are attracted toward the 
critical solution. 

In this note we study equivariant maps of the  SU(2) sigma-model (wave maps)  
from 3+1 dimensional Minkowski spacetime into the three-sphere. 
This system admits a discrete sequence of self-similar solutions and   
has been studied as a toy model for critical collapse behaviour \cite{bizon2}. 
Linear perturbation analysis shows that the "ground state"  is stable, 
while the higher excitations  are unstable. The first excitation acts as 
the critical solution between collapse to the ground state and dispersion. 
Bizo\'n \cite{bizon3} has given a beautiful algorithm how to obtain 
analytically all 
eigenvalues for the linear perturbations of the ground state. However, 
for  the higher excitations, because they are known only numerically, one 
has to rely on numerical tools.

We apply two different numerical methods to calculate  
the first few eigenvalues of the ground state and the first excitation 
i.e. the critical solution.  One method is by shooting and matching  an 
the other by time evolution.   Our analysis shows that the "shooting" 
method, frequently applied for finding the largest  eigenvalues may fail 
for the smaller values.    
 
For the  ground state our results confirm the analytical  results. This is 
important because, as pointed out by Bizo\'n, the eigenvalue problem is not 
standard. Our results show that the analyticity condition for the singular 
boundary value problem chosen by Bizo\'n  are in accordance with the time 
evolution of the system. In addition we find for the critical solution, 
besides of the unstable and the gauge mode,  the first  two stable 
eigenvalues.

\section{The SU(2) $\sigma$-model}
\subsection{Definition of the model}
A smooth mapping $U: (M, g) \rightarrow (S^3, G)$ from a spacetime $M$ with
metric $g$
into the three-sphere is called SU(2) $\sigma$-model if it is a critical point of
the action
\begin{equation}
\label{flataction}
\int_M g^{ab}\partial_a U^A \partial_b U^B G_{AB}.
\end{equation}
Let $M$ be Minkowski spacetime.
We use polar coordinates $(t, r, \theta, \varphi)$ on $M$ and standard coordinates 
$(u, \Theta, \Phi)$ on $S^3$. In these coordinates the metrics are given by
$$g=-dt^2+dr^2+r^2(d\theta^2+\sin^2(\theta) d\varphi^2) $$
\begin{equation}
G=du^2+\sin^2(u)(d\Theta^2+\sin^2(\Theta) d\Phi^2)
\end{equation}
We restrict ourselves to maps of the form 
\begin{equation}
U(t, r, \theta, \varphi)=(u(t, r), \theta, \varphi)
\end{equation}
which are called spherically symmetric and co-rotational.
Under these assumptions the Euler-Lagrange equations associated with the action
(\ref{flataction}) reduce to the single semilinear wave equation
\begin{equation}
\label{sigmamodel}
u_{tt}-u_{rr}-\frac{2}{r}u_r+\frac{\sin(2u)}{r^2}=0.
\end{equation}

\subsection{Self-similar solutions}
A self-similar solution of eq. (\ref{sigmamodel}) is a solution of the form 
\begin{equation}
\label{cssansatz}
u(t, r)=f \left (\frac{r}{T-t} \right )
\end{equation}
for some constant $T>0$.
We substitute the ansatz (\ref{cssansatz}) into eq. (\ref{sigmamodel}) and
obtain
\begin{equation}
\label{csssigmamodel}
f''+\frac{2}{\rho} f'-\frac{\sin(2f)}{\rho^2 (1-\rho^2)}=0
\end{equation}
where $\rho:=\frac{r}{T-t}$ and $':=\frac{d}{d\rho}$.
Note that the domain $\rho \in [0, 1]$ is the backward lightcone of the point 
$(T, 0)$ 
which is called the \emph{singularity} because $u_r(t,0)$ grows unbounded  
for $t \rightarrow T-$.
Therefore $T$ is called the \emph{blow-up time}.
Eq. (\ref{csssigmamodel}) has two singular points ($\rho=0, 1$) and
therefore we have to require the boundary conditions 
\begin{equation}
\label{cssboundary}
f(0)=0 \mbox{ and } f(1)=\frac{\pi}{2}
\end{equation}
to ensure smoothness of solutions.
Bizo\'n \cite{bizon1} has shown existence of a countable family 
$\{ f_0, f_1, \ldots \}$ of smooth solutions of eq. (\ref{csssigmamodel})
which satisfy the boundary conditions (\ref{cssboundary}).
The so-called \emph{ground state} $f_0$ is known in closed form and given by
\begin{equation}
\label{css0}
f_0(\rho)=2 \arctan \rho.
\end{equation}
The \emph{higher excitations} $f_1, f_2, \ldots$ can be constructed numerically
using a standard shooting technique (cf. \cite{bizon2}). 

\subsection{Perturbations}
We are interested in linear perturbations around the solutions $f_0$ and $f_1$
which play a crucial role in dynamical evolution.
We introduce so-called adapted coordinates $(\tau, \rho)$ defined by
\begin{equation}
\label{csscoordinates}
\tau=-\log(T-t) \mbox{ and } \rho=\frac{r}{T-t}.
\end{equation}
These coordinates cover the domain $t<T$ (the singularity is
shifted to $\tau \rightarrow \infty$).
In the new coordinates eq. (\ref{sigmamodel}) becomes
\begin{equation}
\label{sigmamodeltaurho}
u_{\tau \tau}+2\rho u_{\tau \rho}-(1-\rho^2)u_{\rho \rho}+u_\tau -2
\frac{1-\rho^2}{\rho}u_\rho+\frac{\sin(2u)}{\rho^2}=0.
\end{equation}
It is important to note that the class of self-similar solutions 
which blow up as $t \rightarrow T-$ are static (i.e. independent
of $\tau$) in these coordinates.
Inserting
\begin{equation}
u(\tau, \rho)=f_n(\rho)+w(\tau, \rho)
\end{equation}
into eq. (\ref{sigmamodeltaurho}) and neglecting terms of order $w^2$
we obtain a time evolution equation for linear perturbations around the
$n$-th self-similar solution $f_n$.
\begin{equation}
\label{perttimeevol}
w_{\tau \tau}+2\rho w_{\tau \rho}-(1-\rho^2)w_{\rho \rho}+w_\tau -2
\frac{1-\rho^2}{\rho}w_\rho+\frac{2\cos(2f_n)}{\rho^2}w=0
\end{equation}
To ensure regularity we have to require the boundary conditions
\begin{equation}
\label{bcperttimeevol}
w(\tau, 0) \equiv 0 \mbox{ and } w_{\rho \rho}(\tau, 0) \equiv 0.
\end{equation}
We are looking for perturbations of the form
\begin{equation}
\label{eigenmode}
w(\tau, \rho)=e^{\lambda \tau} v_\lambda(\rho)
\end{equation}
where $\lambda$ may be complex in general.
$\lambda$ is said to be an \emph{eigenvalue} and $v_\lambda$ an
\emph{eigenmode} or \emph{eigenfunction}.
Like in the finite-dimensional case one expects that these eigenmodes
determine the dynamics in the neighbourhood of the solution $f_n$ in phase
space.

To obtain an ordinary differential equation for the eigenmodes $v_\lambda$ we
insert ansatz (\ref{eigenmode}) into eq. (\ref{perttimeevol}).
\begin{equation}
\label{odeeigenmode}
v_\lambda''(\rho)+2\frac{(\lambda +1)\rho^2-1}{\rho (\rho+1)(\rho-1)}
v_\lambda'(\rho)+
\frac{\lambda
(\lambda+1)\rho^2+2\cos (2f_n(\rho))}{\rho^2 (\rho+1)(\rho-1)}v_\lambda(\rho)=0.
\end{equation} 
This equation has again two singular points. The regularity conditions are given by
$$ v_\lambda(0)=0 \mbox{, } v_\lambda'(0)=a $$
\begin{equation}
\label{bcodeeigenmode}
v_\lambda(1)=1 \mbox{, } v_\lambda'(1)=-\frac{\lambda^2+\lambda-2}{2\lambda}
\end{equation}
where $a \in \mathbb{R}$ is a free parameter and we have normalized the
eigenfunctions to have $v_\lambda(1)=1$.

\section{Numerical calculation of eigenvalues}
\subsection{Calculating eigenvalues by time evolution}
We consider the Cauchy problem eq. (\ref{perttimeevol}) together with initial
data $w(0,\rho)=f(\rho)$ and $w_\tau(0,\rho)=g(\rho)$ inside the backward
lightcone of the singularity, that is for $\rho \in [0,1]$.
By setting $u_1=w$, $u_2=w_\tau$, $u_3=w_\rho$ and $u=(u_1,u_2,u_3)$ this
problem can be written as a first order system
\begin{equation}
\label{firstorder}
\frac{d}{d\tau} u=L u
\end{equation}
with initial conditions $u_1(0,\rho)=f(\rho)$, $u_2(0,\rho)=g(\rho)$ and $u_3(0,
\rho)=f'(\rho)$.
$L$ is the linear differential operator following from eq.
(\ref{perttimeevol}).
If the Cauchy problem is well-posed one can introduce a one-parameter
family of bounded linear
operators $S(\tau)$ for $\tau \geq 0$ such that the solution to initial data 
$u(0, \cdot)$ is given by
\begin{equation}
\label{solop}
u(\tau, \cdot)=S(\tau)u(0, \cdot).
\end{equation}
We are looking for special solutions defined by eq. (\ref{eigenmode}).
Inserting this ansatz into eq. (\ref{solop})
one obtains
\begin{equation}
\label{eigenmodenv}
S(\tau)(v_\lambda, \lambda v_\lambda, v_\lambda')=
e^{\lambda \tau} (v_\lambda, \lambda v_\lambda, v_\lambda')
\end{equation}
for an eigenmode $v_\lambda$ with eigenvalue $\lambda$.
For fixed $\tau$ we define an inner product by
\begin{equation}
\label{ip}
\langle u(\tau, \cdot) | v(\tau, \cdot) \rangle:=
\int_0^1 \sum_{j=1}^3 u_j(\tau, \rho) \overline{v_j(\tau, \rho)} d\rho
\end{equation}
and a norm by
\begin{equation}
\label{norm}
\|u(\tau, \cdot)\|:=\sqrt{\langle u(\tau, \cdot) | 
u(\tau, \cdot) \rangle}.
\end{equation}
If $v_\lambda$ is an eigenmode with eigenvalue $\lambda$ then we have
\begin{equation}
\|S(\tau)(v_\lambda, \lambda v_\lambda, v_\lambda')\|=
|e^{\lambda \tau}| \|(v_\lambda, \lambda v_\lambda, v_\lambda')\|
=C e^{\mathrm{Re} \lambda \tau}
\end{equation}
where $C$ is constant.
Now we make the following assumptions:
\begin{itemize}
\item The eigenvalues $\{\lambda_0, \lambda_1, \ldots\}$ form a discrete set
\item Arbitrary initial data can be decomposed into a sum of 
eigenmodes
\end{itemize}
Then we can write (real-valued) initial data $u(0, \cdot)$ as
\begin{equation}
\label{eigendecomp}
u(0, \rho)=\sum_{k=0}^\infty (c_k \varphi_k(\rho) +  
\overline{c_k \varphi_k(\rho)})=2 \sum_{k=0}^\infty \mathrm{Re} (c_k
\varphi_k(\rho))  
\end{equation}
where $\varphi_k:=(v_{\lambda_k}, \lambda_k v_{\lambda_k}, v_{\lambda_k}')$ and
$v_{\lambda_k}$ is an eigenmode with eigenvalue $\lambda_k$.

For simplicity we write $\mu_k:=\mathrm{Re}\lambda_k$, $\omega_k:=\mathrm{Im}
\lambda_k$ and without loss of generality we assume that
$\mu_0 \geq \mu_1 \geq \ldots$.
Applying the time evolution operator $S(\tau)$ to initial data $u(0, \cdot)$
yields
\begin{equation}
S(\tau) u(0, \cdot) =2\sum_{k=0}^\infty 
 \mathrm{Re} (c_k e^{\lambda_k \tau} \varphi_k)=
 2\sum_{k=0}^\infty e^{\mu_k \tau} \mathrm{Re}(c_k e^{i \omega_k \tau}
 \varphi_k) .
\end{equation}
So for large $\tau$ and $\mu_0 > \mu_1$ we have
\begin{equation}
\label{g}
\|S(\tau) u(0, \cdot) \| \sim e^{\mu_0 \tau} g(\tau)
\end{equation} 
where $g(\tau):=\|\mathrm{Re}(c_0 e^{i \omega_0 \tau}
 \varphi_0)\|$.
If there are more eigenvalues with the same real part $\mu_0$ then the function 
$g$ in (\ref{g}) is more
complicated but the result is essentially the same.

For computing the real part $\mu_0$ of the dominant eigenvalue one simply
evolves arbitrary initial data and calculates the norm (\ref{norm}). 
As long as the expansion
coefficient $c_0$ in (\ref{eigendecomp}) does not vanish, the logarithm of the 
norm will behave
as $\mu_0 \tau + \log(g(\tau))$ for large $\tau$ and by least-square fitting
one can read off $\mu_0$.
Note that $g$ is constant if and only if $\lambda_0$ is real.

To calculate real parts of other eigenvalues we use the orthogonalization 
method described in \cite{kha}.
Here we only sketch the basic idea. 
Take linearly independent initial data $\phi$ and $\psi$.
Let $P_\phi$ denote the orthogonal projection on $\{\phi\}^\perp$.
One can readily show that
$P_{S(\tau)\phi}S(\tau)P_\phi=P_{S(\tau)\phi}S(\tau)$.
Now we define a third function $\chi:=\psi+\alpha \phi$ where
$\alpha$ is some constant.
Clearly we have $P_\phi \chi=P_\phi \psi$.
Suppose we choose the constant $\alpha$ in such a way that $\chi$ has the
expansion
\begin{equation}
\label{expansion1}
\chi=2 \sum_{k=1}^\infty \mathrm{Re} (c_k \varphi_k).
\end{equation}
Note that the sum starts at $k=1$!
For simplicity we assume that $\lambda_0$ and $\lambda_1$ are real. Then, if 
$\lambda_0 > \lambda_1 >
\mathrm{Re} \lambda_2$, 
\begin{equation}
S(\tau)\chi = 2 e^{\lambda_1 \tau}
 \mathrm{Re}(c_1 \varphi_1) + \dots
\end{equation}
Unfortunately we do not know the right $\alpha$ which is required to obtain
expansion (\ref{expansion1}).
But surprisingly this is not necessary since 
$$ P_{S(\tau)\phi}S(\tau)\psi=P_{S(\tau)\phi}S(\tau)P_\phi \psi=
P_{S(\tau)\phi}S(\tau)P_\phi \chi=$$
\begin{equation}
P_{S(\tau)\phi}S(\tau) \chi = 
e^{\lambda_1 \tau}P_{S(\tau)\phi} \mathrm{Re}(c_1 \varphi_1) + \dots
\end{equation}
Observe that for real $\lambda_0$ we have  
\begin{equation}
P_{S(\tau)\phi} \sim P_{\mathrm{Re} (c_0 \varphi_0)}
\end{equation}
for large $\tau$.
It follows that
\begin{equation}
\|P_{S(\tau)\phi}S(\tau) \psi\| \sim e^{\lambda_1 \tau}
\end{equation}
for large $\tau$ and we can read off $\lambda_1$.
In practical application the orthogonal projection is performed at every
timestep (see \cite{kha} for details).
The probably most important feature of this method is that it 
works with an arbitrary inner product. The
eigenfunctions $\varphi_k$ do not have to be orthogonal.

\subsection{Calculating eigenvalues by shooting}
Another possiblity for calculating eigenvalues is to use eq.
(\ref{odeeigenmode}).
One has to solve an ODE boundary value problem where 
the boundary conditions are given by the regularity requirements
(\ref{bcodeeigenmode}).
This problem can be solved using a standard shooting to a fitting point 
technique.
Although this method is very accurate it has at least two serious disadvantages:
\begin{itemize}
\item It is possible to "miss" some eigenvalues because one has to start with
sufficiently good initial guesses of the shooting parameters.
\item The method cannot be expected to work for large negative eigenvalues (see
\cite{bizon3}).
\end{itemize}
Nevertheless it is a good check for results obtained by time evolution.

\section{Numerical results}
We have calculated four eigenvalues of the
ground state and the first excitation using the orthogonalization method.
The first negative eigenvalue can be obtained by shooting as well.
The positive eigenvalues have been computed before in \cite{bizon2} via 
shooting.
Bizo\'n \cite{bizon3} has also calculated a large number of negative 
eigenvalues for the
ground state using a completely different method.
His results are in good agreement with ours.
Furthermore, in \cite{bizon3} it has been shown analytically that $\lambda=-2$ is an
eigenvalue of the ground state.
 
The eigenvalue $\lambda=1$ corresponds to a so-called \emph{gauge mode}
which is associated with the freedom of choosing the blow-up time $T$ in the
coordinate transformation (\ref{csscoordinates}) (see \cite{bizon2} for a more
detailed explanation). 
This gauge mode has no physical relevance.

We summarize our results in the following two tables.

\begin{table}[h]
\begin{center}
\begin{tabular}{|c|c|c|}
\hline
Eigenmode & Time Evolution & Shooting \\
\hline 
Gauge & 1 & 1 \\
First stable & -0.5424 & -0.54246 \\
Second stable & -2.00 &  \\
Third stable & -3.3 & \\
\hline
\end{tabular}
\end{center}
\caption{Eigenvalues of the ground state}
\end{table}

\begin{table}[h]
\begin{center}
\begin{tabular}{|c|c|c|}
\hline
Eigenmode & Time Evolution & Shooting \\
\hline 
Unstable & 6.3336 & 6.333625 \\
Gauge & 1 & 1 \\
First stable & -0.518 & -0.5186 \\
Second stable & -1.7 &  \\
\hline
\end{tabular}
\end{center}
\caption{Eigenvalues of the first excitation}
\end{table}

Remarkably we do not see any oscillations and therefore all eigenvalues seem
to be real.

\section{Acknowledgments}
We thank Piotr Bizo\'n for helpful discussions. 
This work was supported by the Austrian Fond zur  
F\"orderung der wissenschaftlichen Forschung (FWF) Project P15738
and the Fundacion Federico.

\appendix

\section{Numerical integration of the evolution equation}
\subsection{Integration scheme}
For integrating eq. (\ref{firstorder}) we apply the \emph{second-order 
characteristic method} used e.g. in \cite{gundlach}.
We write eq. (\ref{firstorder}) as
\begin{equation}
\label{system}
\frac{\partial u}{\partial \tau}=A \frac{\partial u}{\partial \rho}+B u
\end{equation}
where $A$ and $B$ are $3 \times 3$ matrices depending on $\rho$.
The canonical decomposition of the matrix $A$ is given by
\begin{equation}
A=\lambda_- A_- + \lambda_+ A_+
\end{equation}
where $\lambda_-$, $\lambda_+$ are the two non-zero eigenvalues of $A$.
$A_-$ and $A_+$ are the orthogonal projections on the eigenspaces.
The (signs of the) eigenvalues of $A$ define the directions of the 
characteristics of the system (\ref{system}).
We discretize the system (\ref{system}) in space by
\begin{equation}
\frac{\partial u}{\partial \tau}= \lambda_- A_- D(\lambda_-) u + \lambda_+ A_+
D(\lambda_+) u +B u
\end{equation}
where $D(\lambda)$ is a finite difference operator producing either 3-point
left-sided derivatives or 3-point right-sided derivatives depending on the sign
of $\lambda$.
For the discretization in time we use a Runge-Kutta-like scheme (cf.
\cite{gundlach}).

\subsection{Boundary conditions}
At the boundary $\rho=1$ we do not have to impose any boundary condition because
it turns out that $\lambda_\pm \leq 0$ for $\rho \geq 1$.
Therefore we do not need right-sided derivatives here.
Physically this is clear since $\rho=1$ is the past lightcone of the singularity
and therefore no information can come in from outside.
At $\rho=1-\Delta \rho$ (where $\Delta \rho$ denotes the mesh of our grid) 
we do need a right-sided derivative and since we are
using 3-point approximations we need one point outside the lightcone $\rho=1$.
Therefore we integrate the equation on $\rho \in [0, 1+\Delta \rho]$.

At the center $\rho=0$ we have the regularity conditions (cf.
(\ref{bcperttimeevol}))
\begin{equation}
u_1(\tau, 0) \equiv 0 \mbox{ and } \frac{\partial u_3}{\partial \rho}(\tau, 0)
\equiv 0.
\end{equation} 
To impose these boundary conditions we define ghost points by
\begin{equation}
u_1(\tau, -\rho)=-u_1(\tau, \rho)
\end{equation}
and use the same integration scheme as in the interior.


\begin{thebibliography}{10}
\bibitem{bizon1} P. Bizo\'n, \emph{Equivariant self-similar wave maps from 
Minkowski spacetime into the 3-sphere}, Comm. Math. Phys. \textbf{215}, 45
(2000)
\bibitem{bizon2} P. Bizo\'n, T. Chmaj and Z. Tabor, 
\emph{Dispersion and collapse of wave maps}, Nonlinearity \textbf{13}
(2000) 1411--1423
\bibitem{kha} T. Hara, T. Koike and S. Adachi, \emph{Renormalization group and
critical behaviour in gravitational collapse}, gr-qc/9607010
\bibitem{bizon3} P. Bizo\'n, \emph{An unusual eigenvalue problem},
Acta Phys. Polon. 36, 5 (2005)
\bibitem{gundlach} J. M. Mart\'in-Garc\'ia and C. Gundlach, 
\emph{All nonspherical perturbations of the Choptuik spacetime decay},
Phys. Rev. \textbf{D59} (1999) 064031
\end{thebibliography}
\end{document}